\documentclass[a4paper,12pt]{article}
\usepackage{feynmp-auto}
\usepackage{amsmath, amsfonts, amssymb}
\usepackage{graphicx}
\usepackage{enumerate}
\usepackage{hyperref}
\usepackage{latexsym}
\usepackage{dsfont}
\usepackage{hepnicenames}
\usepackage{enumerate}
\usepackage{mathrsfs,graphicx,rotating,mathtools,booktabs}
\usepackage{hyperref}\usepackage{slashed}
\usepackage[nosort]{cite}
\usepackage[table,xcdraw,dvipsnames]{xcolor}
\usepackage{youngtab}
\usepackage{bm}
\usepackage{graphicx}
\usepackage{multirow,multicol}
\hypersetup{colorlinks,bookmarksopen,bookmarksnumbered,
linkcolor=blus,pdfstartview=FitH,urlcolor=rossos,citecolor=verde}
\allowdisplaybreaks

\renewcommand{\[}{\left[}
\renewcommand{\]}{\right]}
\renewcommand{\(}{\left(}
\renewcommand{\)}{\right)}

\newcommand{\rmd}{\mathrm{d}}
\newcommand{\al}{\alpha}

\newcommand{\gam}{\gamma}

\newcommand{\sig}{\sigma}

\newcommand{\Om}{\Omega}

\newcommand{\mio}[1]{}

\def\bpm{\begin{pmatrix}}
\def\epm{\end{pmatrix}}

\usepackage{mathrsfs}

\newcommand{\sfrac}[2]{#1/#2}

\allowdisplaybreaks
\usepackage{multicol}
\usepackage{color}
\definecolor{rosso}{cmyk}{0,1,1,0.4}
\definecolor{rossos}{cmyk}{0,1,1,0.55}
\definecolor{rossoc}{cmyk}{0,1,1,0.2}
\definecolor{blu}{cmyk}{1,1,0,0.3}
\definecolor{blus}{cmyk}{1,1,0,0.6}
\definecolor{bluc}{cmyk}{1,1,0,0.1}
\definecolor{verde}{cmyk}{0.92,0,0.59,0.25}
\definecolor{verdec}{cmyk}{0.92,0,0.59,0.15}
\definecolor{verdes}{cmyk}{0.92,0,0.59,0.4}

\newcommand{\eq}[1]{~{\rm (\ref{eq:#1})}}

\newcommand{\s}{\,{\rm s}}

\newcommand{\GeV}{\,{\rm GeV}}

\newcommand{\cm}{\,{\rm cm}}

\def\circa#1{\,\raise.3ex\hbox{$#1$\kern-.75em\lower1ex\hbox{$\sim$}}\,}

\newcommand{\beq}{\begin{equation}}
\newcommand{\eeq}{\end{equation}}

\newcommand{\bea}{\begin{eqnarray}}
\newcommand{\eea}{\end{eqnarray}}
\newcommand{\be}{\begin{equation}}
\newcommand{\ee}{\end{equation}}
\font\tenrsfs=rsfs10 at 12pt
\font\sevenrsfs=rsfs7
\font\fiversfs=rsfs5
\newfam\rsfsfam
\textfont\rsfsfam=\tenrsfs
\scriptfont\rsfsfam=\sevenrsfs
\scriptscriptfont\rsfsfam=\fiversfs

\newsavebox\MBox

\def\circa#1{\,\raise.3ex\hbox{$#1$\kern-.75em\lower1ex\hbox{$\sim$}}\,}
\makeatletter

\font\ital=cmu10

\def\hhref#1{\href{http://arxiv.org/abs/#1}{arXiv:#1}}
\usepackage{xstring}
\newcommand{\hhrefq}[1]{\IfSubStr{#1}{:}{\href{http://inspirehep.net/search?ln=en&ln=en&p=#1&of=hb&action_search=Search&sf=&so=d&rm=&rg=25&sc=0}{InSpires:#1}}{\hhref{#1}}}

\def\art{\@ifnextchar[{\eart}{\oart}}
\def\eart[#1]#2#3#4#5#6{{\rm #2}, {\em #3 \bf #4} {\rm (#6) #5} ({\em #1})}
\def\article{\@ifnextchar[{\earticle}{\oarticle}}
\def\oarticle#1#2#3#4#5#6{{\rm #1}, {\ital ``#6''}, {\rm #2 #3 (#5) #4}}
\def\earticle[#1]#2#3#4#5#6#7{{\rm #2}, {\ital ``#7''}, {\rm #3 #4 (#6) #5}  [\hhrefq{#1}]}
\def\hepart[#1]#2{{\rm #2, \sl#1}}
\def\heparticle[#1]#2#3{#2, {\ital ``#3''} [\hhrefq{#1}]}
\newcommand{\doi}[1]{\href{http://dx.doi.org/#1}{[link]}}

\renewenvironment{thebibliography}[1]
     {\begin{multicols}{2}[\section*{\refname}]%
      \@mkboth{\MakeUppercase\refname}{\MakeUppercase\refname}%
      \list{\@biblabel{\@arabic\c@enumiv}}%
           {\settowidth\labelwidth{\@biblabel{#1}}%
            \leftmargin\labelwidth
            \advance\leftmargin\labelsep
            \@openbib@code
            \usecounter{enumiv}%
            \let\p@enumiv\@empty
            \renewcommand\theenumiv{\@arabic\c@enumiv}}%
      \sloppy
      \clubpenalty4000
      \@clubpenalty \clubpenalty
      \widowpenalty4000%
      \sfcode`\.\@m}
     {\def\@noitemerr
       {\@latex@warning{Empty `thebibliography' environment}}%
      \endlist\end{multicols}}

\newcounter{alphaequation}[equation]
\def\thealphaequation{\theequation\hbox to
0.6em{\hfil\alph{alphaequation}\hfil}}
\def\eqnsystem#1{
\def\@eqnnum{{\rm (\thealphaequation)}}
\def\@@eqncr{\let\@tempa\relax \ifcase\@eqcnt \def\@tempa{& & &} \or
  \def\@tempa{& &}\or \def\@tempa{&}\fi\@tempa
  \if@eqnsw\@eqnnum\refstepcounter{alphaequation}\fi
\global\@eqnswtrue\global\@eqcnt=0\cr}
\refstepcounter{equation} \let\@currentlabel\theequation \def\@tempb{#1}
\ifx\@tempb\empty\else\label{#1}\fi
\refstepcounter{alphaequation}
\let\@currentlabel\thealphaequation
\global\@eqnswtrue\global\@eqcnt=0 \tabskip\@centering\let\\=\@eqncr
$$\halign to \displaywidth\bgroup \@eqnsel\hskip\@centering
$\displaystyle\tabskip\z@{##}$&\global\@eqcnt\@ne
\hskip2\arraycolsep\hfil${##}$\hfil& \global\@eqcnt\tw@\hskip2\arraycolsep
$\displaystyle\tabskip\z@{##}$\hfil
\tabskip\@centering&\llap{##}\tabskip\z@\cr}
\def\endeqnsystem{\@@eqncr\egroup$$\global\@ignoretrue} \makeatother

\definecolor{Gray}{gray}{0.95}

\def\bal#1\eal{\begin{align}#1\end{align}}

\graphicspath{{./}}

\begin{document}

{CERN-TH-2018-052\hfill IFUP-TH/2018}

\vspace{1.5cm}

\begin{center}
{\Large\LARGE \bf \color{rossos}
Bounds on Dark Matter \\[3mm] annihilations from 21 cm data}\\[1cm]
\textbf{Guido D'Amico$^a$, Paolo Panci$^a$, Alessandro Strumia$^{a,b,c}$}\\[7mm]

{\it $^a$ CERN, Theory Division, Geneva, Switzerland}\\[1mm]
{\it $^b$ Dipartimento di Fisica dell'Universit{\`a} di Pisa}\\[1mm]
{\it $^c$ INFN, Sezione di Pisa, Italy}\\[1mm]

\vspace{0.5cm}

{\large\bf\color{blus} Abstract}
\begin{quote}\large
The observation of an absorption feature in the 21 cm spectrum
at redshift $z\approx 17$ implies bounds on Dark Matter annihilations for a broad range of masses, given that significant heating of the intergalactic medium would have erased such feature.
The resulting bounds on the DM annihilation cross sections are comparable to the strongest ones from all other observables.
\end{quote}

\thispagestyle{empty}
\bigskip

\end{center}

%\tableofcontents

\setcounter{footnote}{0}

\section{Introduction}
The EDGES experiment recently reported the first measurement of the global 21-cm spectrum~\cite{EDGES}, which is an observable sensitive to the temperature of the gas at redshift $z\approx 17$.
This allows to constrain the Dark Matter (DM) annihilation cross section, as the annihilation products would heat the gas~\cite{astro-ph/0603425,astro-ph/0606483,0907.0719,1408.1109,1603.06795}.

The signal seen by EDGES is the absorption of light at energy equal to $\Delta E %= 5.87\,{\mu}{\rm eV}
=0.068\,{\rm K}= 2\pi/(21 \cm)$ in the rest frame of the gas.
This is the energy difference between the ground states of hydrogen with total spin $S=0$ or $1$ (depending on the relative spin between electron and proton).
Cosmological red-shifting brings the signal to radio frequencies of order $\sim 100\, \mathrm{Mhz}$.
The signal is reported in terms of the average of the difference between the brightness temperature and the one of the background radiation, given by~\cite{Zaldarriaga:2003du,astro-ph/0608032}
\beq
T_{21}(z) \approx  23 \,   \mathrm{mK}
%(1 + \del)
\left(1 - \frac{T_\gamma(z)}{T_{S}(z)} \right) \( \frac{\Om_b h^2}{0.02} \) \(\frac{0.15}{\Om_m h^2} \)^{1/2}
\sqrt{\frac{1+z}{10}} \, x_{\rm HI} .
\eeq
Here $x_{\rm HI}$ is the number fraction of neutral hydrogen, very close to 1.
Next, $T_\gamma(z)$ is the background photon temperature, expected to be dominated by the low-energy tail of CMB photons, so that $T_\gamma = T_{\rm CMB}= 2.7\,{\rm K}(1+z)$.
$T_S $ is the `spin-temperature', which defines the relative population of the two spin levels of hydrogen ground state as $n_1/n_0 \equiv 3 e^{-\Delta E/T_S}$.

\smallskip

According to standard cosmology, the gas, composed mainly of neutral hydrogen, thermally decoupled from CMB at $z\approx 150$.
After thermal decoupling, the gas cools like any non-relativistic particle, such that $T_{\rm gas}/T_{\rm CMB}\propto (1+z)$.
%, namely $T_{\rm gas} \approx 9.3\,{\rm K}$ at $z=20$.
At $z \circa{<} 20$ the first light from stars (re)couples the 21 cm two-state system to gas, such that its spin temperature becomes $T_S=T_{\rm gas}$.
This is a sensible assumption, in the limit of a large Ly-$\al$ radiation rate, and no heating of the gas due to $X$-ray radiation from first stars.
In any case, using detailed balance, the spin temperature has to be higher than $T_{\rm gas}$, as any other source of radiation is hotter.
At even lower redshifts, $z\circa{<}15$, star-light heats the gas to temperatures higher than the CMB, and the $T_{21}$ signal goes to zero.
EDGES measured an absorption feature centered at a frequency of $\approx 78 \, \mathrm{MHz}$, translating to a redshift $z = 17.2$,
%$16\circa{<}z\circa{<} 19$,
at which
\beq
T_{21} \approx -500^{+200}_{-500} \, \mathrm{mK} \quad \textrm{(99\% C.L.)}. % \pm 70 \, \rm mK.
\label{eq:EDGES}
\eeq
The expectation from standard astrophysics with non-interacting DM is $T_{21}\approx - 200\,{\rm mK}$.
Thereby, the gas temperature inferred from eq.\eq{EDGES}
is about a factor of 2 lower than what expected. The statistical significance of the anomaly is claimed to be 3.8$\sigma$.
This could be due to systematic issues, or to astrophysical processes increasing $T_\gamma$~\cite{1802.07432,1803.01815}, or maybe to new physics~\cite{Barkana,1802.10577,1802.10094}.
In our paper, we will not address the possible origin of the anomaly.
Rather, we use the fact that an absorption feature is observed to set bounds on DM annihilations.

\section{Bound on DM annihilations}
DM annihilation products can considerably heat the gas, therefore suppressing the observed absorption feature, even erasing it if DM heating is too large.
% If the $3.8\sigma$ anomaly stands, the bound on the gas temperature would be 3.8 stronger than the claimed uncertainty.
% We do not follow this path.
To give bounds on DM annihilations, we will not rely on the actual value of the strong absorption signal, but we conservatively impose that DM heating does not erase the absorption feature observed down to $z\approx 16$.

DM annihilations will heat the gas in two ways.
First, DM annihilations around the period of thermal decoupling from the CMB can increase the amount of free electrons above the value predicted by the Standard Model, $x_e = n_e/n_b\approx 2~\times~10^{-4}$.
A higher $x_e$ delays hydrogen/CMB decoupling, increasing $T_{\rm gas}$ at lower redshifts since the gas has less time to cool adiabatically.
More importantly, DM annihilations directly heat the hydrogen gas through energy injection, increasing $T_{\rm gas}$.
A higher $T_{\rm gas}$ will result in a modification of the $T_{21}$ spectrum~\cite{1209.2120,1408.1109,1603.06795}.

%In setting bounds, we conservatively neglect DM over-densities.
%This should be a good approximation, given that
%Later, star-light heated $T_{\rm gas}$.

In the presence of DM annihilations, the temperature of the gas $T_{\rm gas}$ and the free electron fraction $x_e$ evolve as dictated by
\begin{eqnsystem}{sys:eqs}
    \frac{{\rm d}x_e}{{\rm d}z} &=& \frac{\mathcal{P}_2}{(1+z)H(z)}  \[ \mathcal \alpha_H(T_{\rm gas})  n_{\rm H} x_e^2 - \mathcal \beta_H(T_{\rm gas}) e^{-E_\alpha/T_{\rm gas}} (1-x_e) \]+ \\  \nonumber
    &&- \frac{1}{(1+z) H(z)} \left. \frac{\rmd E}{\rmd V \rmd t} \right|_{\rm inj} \frac{1}{n_H}
    \(\frac{f_{\rm ion}(z)}{E_0}+\frac{(1-\mathcal{P}_2) f_{\rm exc}(z)}{E_\al} \) \, ,\\
    \frac{{\rm d}T_{\rm gas}}{{\rm d}z} &=& \frac{1}{1+z} \[ 2 T_{\rm gas} - \gam_{\rm C} \(T_\gam(z) - T_{\rm gas} \) \]+ \\ \nonumber
    &&- \frac{1}{(1+z) H(z)} \left. \frac{\rmd E}{\rmd V \rmd t} \right|_{\rm inj} \frac{1}{n_H} \frac{2 f_{\rm heat}(z)}{3 (1+x_e+f_{\rm He})}.
\label{eq:evolAnn}
%    \frac{{\rm d}x_e}{{\rm d}z} &=& \frac{\mathcal P_2}{(1+z)H(z)}  \[(\mathcal \alpha_H(T_{\rm gas})  n_{\rm H} x_e^2 - \mathcal \beta_H(T_{\rm gas}) e^{-E_\alpha/T_{\rm gas}} (1-x_e) \] \, , \\
%    \frac{{\rm d}T_{\rm g}}{{\rm d}z} &=& \frac{1}{1+z} \[ 2 T_{\rm gas} - \gam_{\rm C} \(T_\gam(z) - T_{\rm gas} \) \] \, ,
%\label{eq:evol}
\end{eqnsystem}
The upper line in each equation describes standard cosmology:
$E_{\alpha} = 3 E_0/4$ is the Lyman-$\al$ energy, and $E_0 = 13.6 \rm eV$ is the binding energy of hydrogen in its ground state,
 % $B(z) = \med{\rho^2}/\med{\rho}^2$ is the boost factor discussed below,
$\beta_H$ is the effective photoionization rate for an atom in the $2s$ state, and $\alpha_H$ is the case-B recombination coefficient.
We defined the dimensionless coefficient
\beq
    \gamma_{\rm C} \equiv \frac{8 \sig_{\rm T} a_r T_\gam^4}{3 H m_e c} \frac{x_e}{1 + f_{\rm He} + x_e} \, ,
    \label{eq:compton}
\eeq
where $\sig_{\rm T}$ is the Thomson cross-section, $a_r$ the radiation constant, $m_e$ the electron mass and $f_{\rm He}$ the number fraction of helium.
The coefficient $\mathcal{P}_2$ represents the probability for an electron in the $n = 2$ state to get to the ground state before being ionized, given by~\cite{Giesen:2012rp}
\beq
    \mathcal{P}_2 = \frac{1 + K_H \Lambda_H n_H (1-x_e)}{1 + K_H (\Lambda_H + \beta_H) n_H (1-x_e)} \, ,
\eeq
where $\Lambda_H = 8.22 \, {\rm s^{-1}}$ is the decay rate of the $2s$ level, and the factor $K_H = \pi^2 /(E_\al^3 H(z))$ accounts for the cosmological redshifting of the Ly-$\alpha$ photons.
We solve the above equations starting from an initial redshift $z_M$ before recombination, imposing $x_e(z_M)=1$ and $T_{\rm gas}(z_M)=T_{\rm CMB}(z_M)$.
We use $z_M=1400$ and we have checked that solutions do not change using a different starting point.

The lower terms in equations~(\ref{sys:eqs}) describe the additional effect of DM annihilations.
The energy injection rate per unit volume due to DM is
\be
\label{eq:Einj}
\left. \frac{\rmd E}{\rmd V \rmd t} \right|_{\rm inj}
= \rho_{\rm DM}^2 f_{\rm DM}^2 \frac{\langle \sig v \rangle }{M_{\rm DM}} \, ,
\ee
with $f_{\rm DM}$ the fraction of the dark matter which annihilates.
The dimensionless factors $f_{\rm c}(z)$ take into account the efficiency of deposition in the gas of the injected energy in three different channels c, namely ionization (ion), excitation (exc), and heating (heat), as defined in~\cite{1506.03811, 1506.03812}.
In our calculations, we computed them according to~\cite{1506.03812}.
The $f_{\rm c}(z)$ depend on the primary annihilation channel and on the DM mass, and take into account the delay between the injection and the deposition of energy.
An important ingredient which needs to be considered at low redshifts is the effect of structure formation, which enhances the injected energy due to the DM annihilation with respect to the smooth background.
This can be estimated by replacing, in eq.~\eqref{eq:Einj}, $\rho_{\rm DM}^2 \to \langle \rho_{\rm DM}^2 \rangle = B(z) \langle {\rho}_{\rm DM}\rangle^2$.
The boost factor $B(z)$ is the variance of the DM power spectrum, which is subject to considerable uncertainty (see for instance figure 2 of~\cite{1604.02457}).
Since the $f_{\rm c}(z)$ functions depend on the history of the energy injection at redshifts previous to $z$,
they will  involve the time integral of the boost factor.

% It varies slowly with redshift, and equals $f_{\rm eff}\approx 0.15-0.25$  for $M_{\rm DM}\circa{>} 10\GeV$.
% A larger $f_{\rm eff}\approx 0.4-0.6$ arises if DM annihilates mostly into $e^\pm$ or $\gamma$.
% A smaller  $f_{\rm eff}$ arises if DM annihilates mostly into neutrinos.

%Finally, $B\ge1 $ is the cosmological boost factor which accounts for DM inhomogeneities.
%This is better known than analogous astrophysical boost factors.
In our results, we show constraints for two different boost factors.
A conservative choice, denoted as ``Boost 1'' in our plots, is the smallest boost factor from~\cite{1408.1109},
\beq
B(z) \approx 1+ \frac{1.6~\times~10^5}{(1+z)^{1.54}}\,{\rm erfc}\(\frac{1+z}{20.5}\) \, ,
\label{eq:Boost1}
\eeq
which evaluates to $B \approx 217$ at $z\approx 20$ and roughly
agrees with the smallest boost factor in fig.~19 of~\cite{1012.4515}.
Higher boost factors are considered in the literature.
To illustrate the effect that the boost factor has on the constraints, we also plot results for a less conservative choice,
denoted as ``Boost 2'', obtained from
a halo model calculation, considering an Einasto profile with substructures and minimum halo mass of $10^{-6} M_{\odot}$ (figure 2 of~\cite{1604.02457}).

We notice that the use of the cosmological boost factor is justified for the $T_{21}$ observable at hand.
Indeed the first starlight induces a 21 cm signal from roughly all baryons in the universe, not only from those in over-dense regions close to structures that contain the first stars, as $X$-ray photons lead to a roughly uniform radiation flux.
DM annihilations dominantly happen in many small overdensities, enhanced by the boost factor, but annihilation products produce a broad spectrum of radiation which will spread the heat leading to a roughly uniform heating~\cite{1408.1109}.

\begin{figure}[pht]
    \centering
    \includegraphics[width=0.475\textwidth]{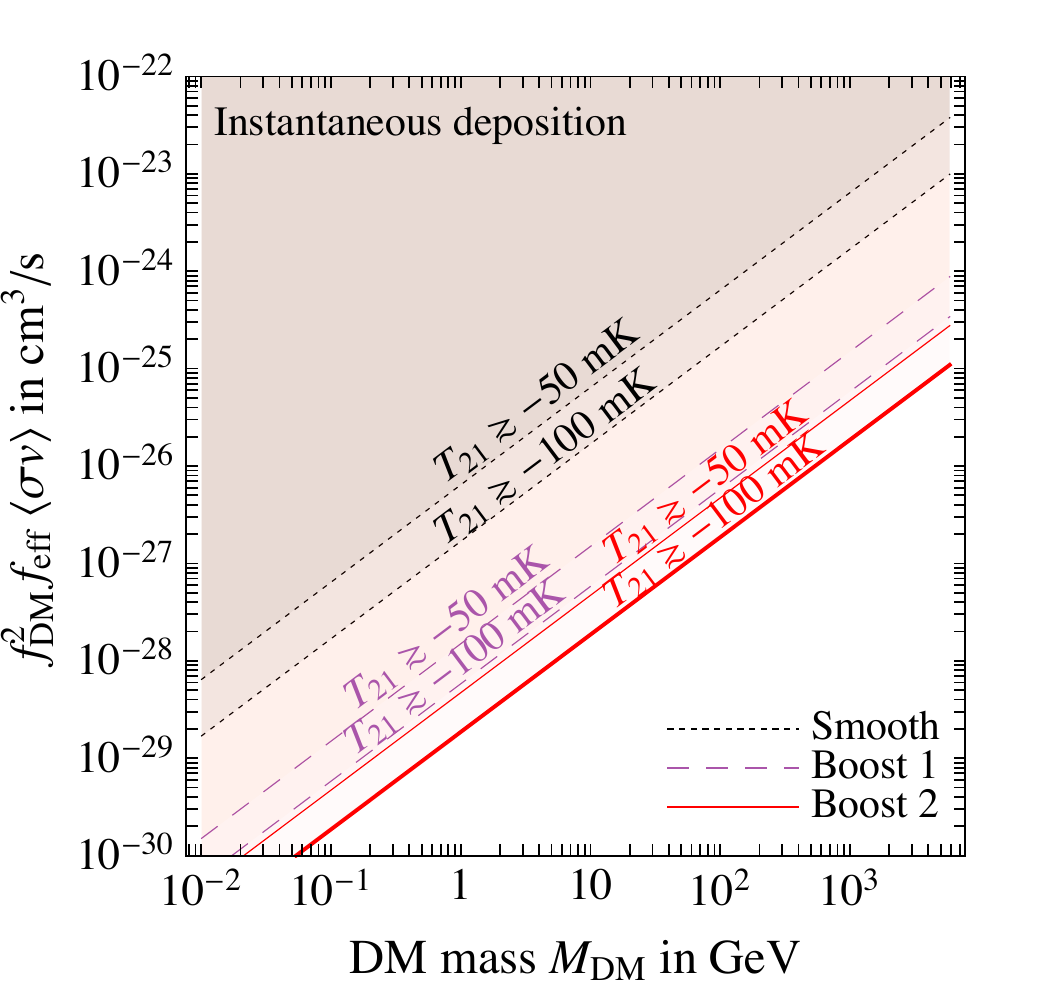}\
    \includegraphics[width=0.475\textwidth]{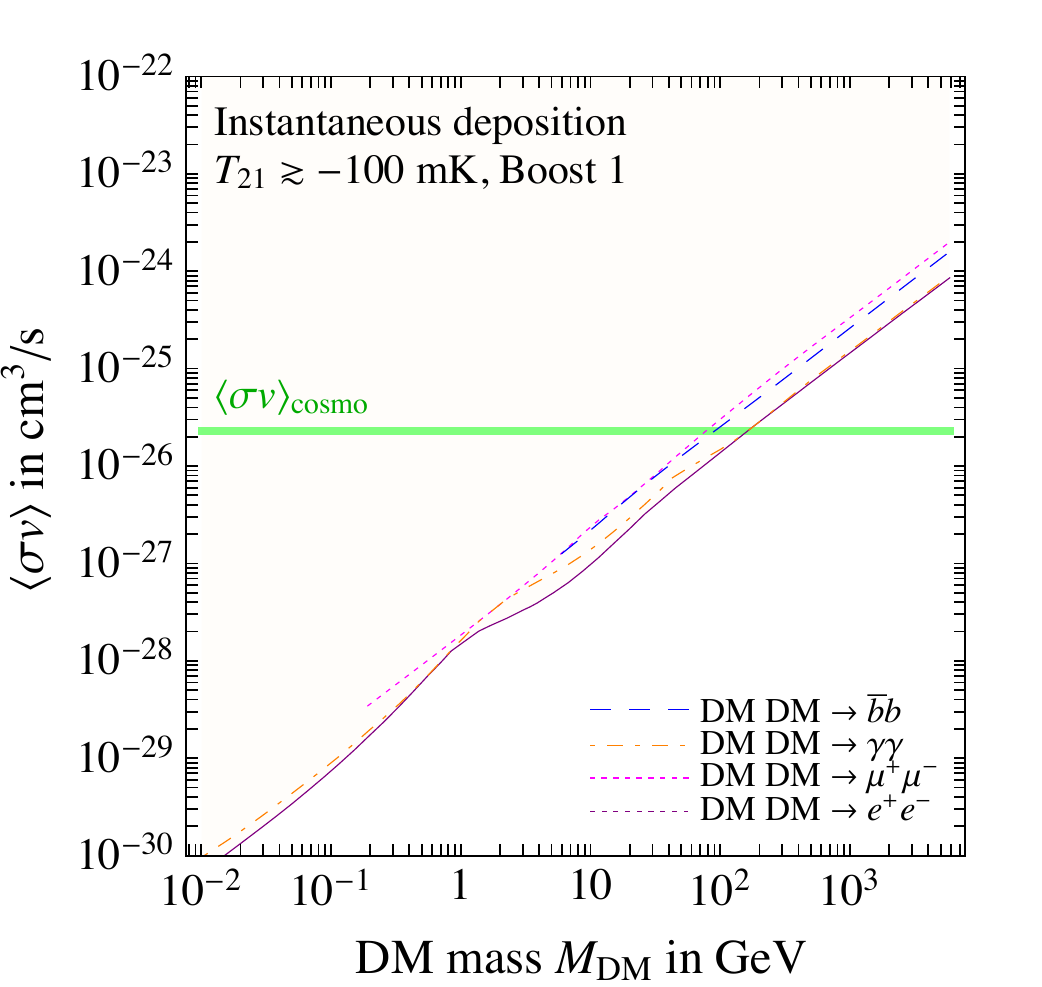}\
    \includegraphics[width=0.475\textwidth]{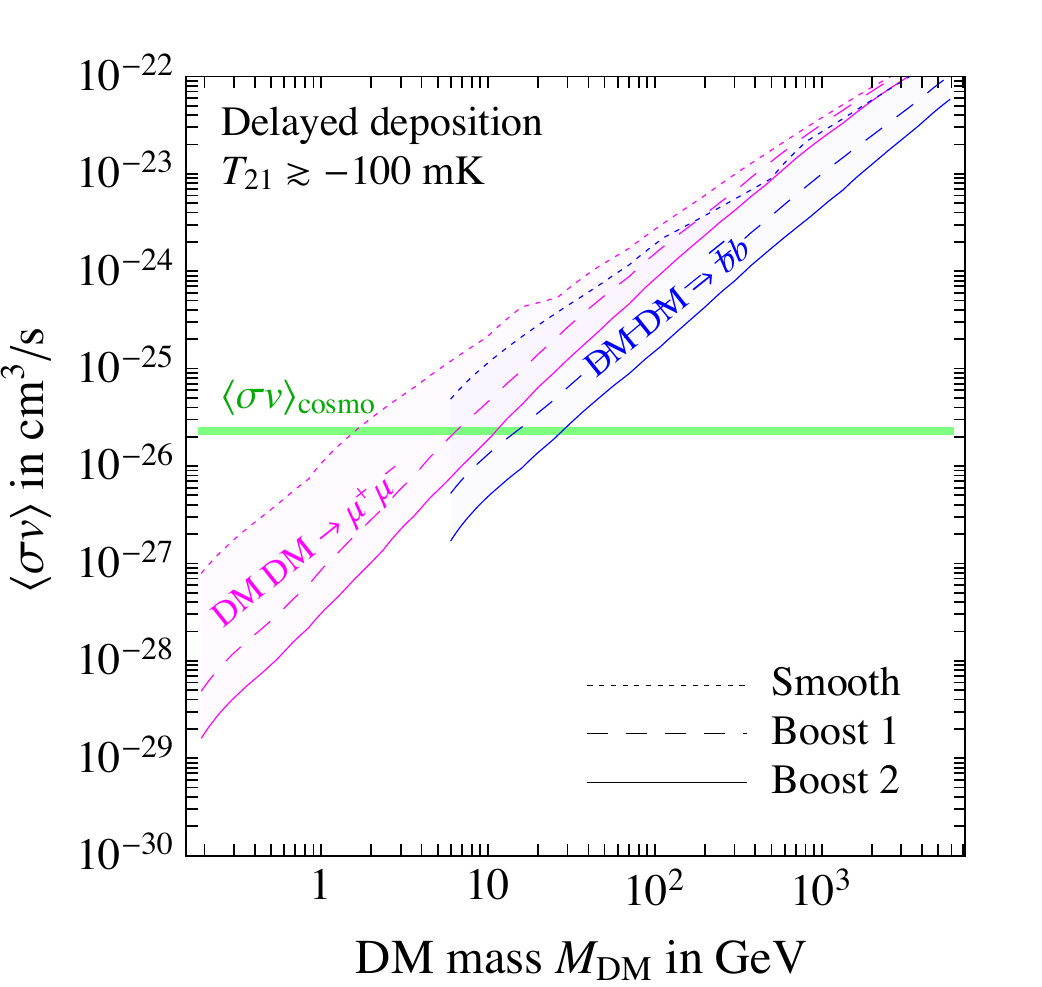}\
    \includegraphics[width=0.475\textwidth]{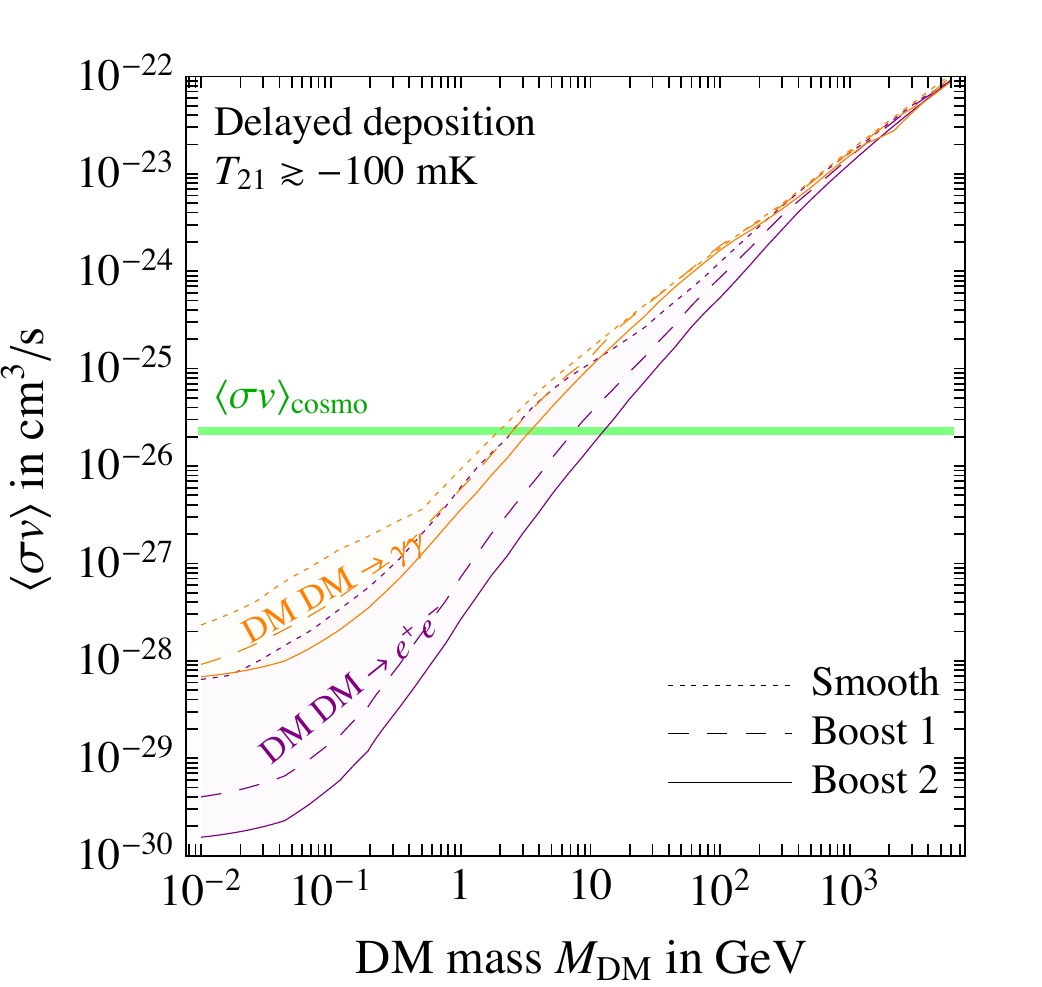}\
    \caption{\em \label{fig:fsigmav} {\bf Upper row:}
    Bounds on DM annihilation cross sections in the instantaneous deposition approximation.
    The left panel shows bounds on the cross section times efficiency factor $f_{\rm eff}$ and fraction of annihilating DM $f_{\rm DM}^2$, obtained by demanding that the $21\cm$ absorption feature  is not depleted from the value of standard cosmology ($-200\,{\rm mK}$ in our computation) down to $-100$ or $-50\,{\rm mK}$ because of DM heating.
    We take into account two different cosmological boost factors, and also show the weaker bound obtained by (irrealistically) ignoring DM clustering.
    The right panel shows bound on the cross section for a few main annihilation channels (bottoms, photons, muons and electrons), using the $f_{\rm eff}$ described in the text, the mild boost factor of eq.~\eqref{eq:Boost1} and demanding that $T_{21} \gtrsim 100 \, \mathrm{mK}$.
    %Here we assume that DM is dominated by a single component, so that $f_{\rm DM}=1$. If $f_{\rm DM}^2 < 1$, the bounds simply rescale.
    {\bf Bottom row:} Bounds on DM annihilation cross sections using delayed energy deposition and numerical primary spectra as described in the text. We demand that $T_{21} \gtrsim 100 \, \mathrm{mK}$, showing the results for two different boost factors, as well as ignoring DM clustering.
    The left (right) panel shows DM decaying into bottom quarks or muons (photons or electrons).}
\end{figure}

\smallskip

In fig.~\ref{fig:fsigmav} we show the constraints on DM annihilations obtained by imposing that the DM correction  to  $T_{21}$ does not suppress by more than a factor of 2 or 4 the $T_{21}$ resulting from standard astrophysics, close to $- 200\,{\rm mK}$ as inferred by solving eqs.~(\ref{sys:eqs}) without the DM contribution.

We show bounds for a few different cases, as follows.
In the upper row of fig.~\ref{fig:fsigmav} we consider the instantaneous deposition approximation.
This means that we assume that a fraction $f_{\rm eff}$ of the energy produced by DM annihilation at some redshift is immediately transferred to the plasma, using a simplified approach (``SSCK'' approximation) proposed in~\cite{astro-ph/0310473}, based on earlier work by~\cite{Shull:1982zz}:
\be
f^{\rm SSCK}_{\rm ion} = f^{\rm SSCK}_{\rm exc} = f_{\rm eff} \frac{1 - x_e}{3} \, , \qquad
f^{\rm SSCK}_{\rm heat} = f_{\rm eff} \frac{1 + 2 x_e}{3} \, .
\ee
The upper left panel shows constraints on $f_{\rm eff} \langle \sigma v \rangle$.
It shows the effects of the boost factor, for two different choices of observed $T_{21}$, in the instantaneous deposition approximation (with the SSCK prescription).
These  bounds are well approximated by:
\beq
f_{\rm DM}^2 f_{\rm eff} \langle \sigma v \rangle
< 10^{-26} \frac{\cm^3}{\s} \frac{M_{\rm DM}}{100 \GeV}\times
\left\{\begin{array}{ll}
0.62 &  \hbox{imposing $ T_{21} \circa{>} -100 \,{\rm mK}$}\\
1.57 &\hbox{imposing $ T_{21} \circa{>} -50 \,{\rm mK}$}\\
\end{array}\right. \ .
\label{eq:bounds}
\eeq
%$f_{\rm eff}$ depends on the DM mass and annihilation products.
In the upper right panel we specialize the constraints to a few representative channels: electrons, muons, photons and bottom quarks.
%{\color{red} In a way, this can be regarded as a definition of $f_{\rm eff}$, when one considers a specific decay mode of dark matter.}
We derive the bounds by rescaling eq.~\eqref{eq:bounds} with the $f_{\rm eff}$ from~\cite{1506.03811}, extended to lower energies using the formulae{} and the numerical results given in~\cite{1506.03811,1506.03812}.

\smallskip

The bottom row of figure~\ref{fig:fsigmav} shows again the bounds for some representative DM annihilation channels, but considering a full calculation by convolving the primary spectra provided in~\cite{1012.4515} with the delayed transfer functions of~\cite{1506.03811}.
The effects due to the boost factor vary with DM mass and annihilation channel.
For DM particles annihilating directly into photons or electrons the boost has little effect on the bounds at high DM masses.
This happens because energetic photons and electrons deposit in the gas a relevant amount of their energy only after some time.
%so that those produced after structure formations do not have time to interact with the gas, influencing the 21 cm structure.
In particular for photons the effect is quite small for the full range of mass we consider.
Physically, this can be understood by the fact that the instantaneous deposition approximation becomes poor for highly energetic particles, which were either produced at a redshift in which structures were not already formed and interact with the gas only later, or do not have time to interact with the gas if produced when the boost enhancement becomes important.
On the other hand, energy deposition is well approximated as instantaneous for primary annihilation channels (such as quarks, $\tau$, $W$, $Z$ and $h$) with a broad low-energy spectrum of secondary products, and for primary muons and electrons injected at low energies.
%In these cases, for masses below $1-10 \, \mathrm{GeV}$, the instantaneous deposition is a more reasonable approximation.
The small discrepancy at low masses between the upper right panel and the correct bounds of the bottom row can be attributed to the fact that the $f_{\rm eff}$ derived in~\cite{1506.03811} is the effective deposited fraction relevant for CMB bounds, while here we are interested in different physics.

The 21 cm bounds are comparable to bounds from the CMB (which rely on global fits which assume standard cosmology)~\cite{1506.03811,1502.01589}, and to bounds from indirect detection searches (subject to astrophysical uncertainties)~\cite{1503.02641}.
With respect to the latter case, our bounds apply to a broader range of DM masses.

\section{Conclusions}
We derived strong bounds on DM annihilation cross-sections by demanding that heating due to the annihilations does not erase the 21 cm  absorption feature observed from sources around $z \approx 17.2$.
Even imposing this conservative view, adopting a quite mild cosmological boost factor, DM with an $s$-wave cross-section that reproduces the cosmological DM abundance, $\langle \sigma v\rangle \approx 2.3~\times~10^{-26}\cm^3/\sec$, is excluded for DM masses $M_{\rm DM} \circa{<}3-30\GeV$, depending on the annihilation channel.

The fact that the 21 cm absorption feature seems anomalously stronger than what expected on the basis of collisionless DM is receiving large attention.
In particular, a large baryon/DM interaction in special models of DM with a subleading millicharged component has been immediately considered~\cite{Barkana,1802.10577,1802.10094} as an explanation for the cooling.
We would like to stress here that this explanation, if valid at all~\cite{1803.02804,1803.03091,1803.03245}, comes from an incomplete analysis which neglects the heating caused by DM annihilation.
In these models, in the limit where $T_{\rm DM}\ll T_{\rm gas}$ and where the two components interact strongly enough that they re-thermalise, the gas temperature is reduced at most by a factor $ \sfrac{T'_{\rm gas}}{T_{\rm gas}} =\sfrac{n_b}{(n_b + n_{\rm DM})} $, such that a DM lighter than a few GeV is needed to fully explain the anomaly.
In our analysis, we point out that annihilations of such a light DM are strongly constrained, as it can inject electrons and low-energy photons, which could heat the gas more than it is cooled.
More in general, a large class of models which posit a DM/baryon interaction will feature DM annihilation, whose energy injection must be taken into account.

\appendix

\footnotesize
\bibliographystyle{abbrv}

\subsubsection*{Acknowledgments}
This work was supported by the ERC grant NEO-NAT. We thank K.~Blum and A.~Ferrara for discussions.
We are particularly grateful to T.~Slatyer for helpful discussions and for emphasizing to us the importance of the delayed energy deposition for highly energetic electrons and photons~\cite{1803.09739}.

\end{document}